\documentclass[aps,prb,twocolumn,showpacs,showkeys,preprintnumbers,groupedaddress,amsmath,amssymb,showpacs]{revtex4-1}

\usepackage{graphicx}
\usepackage{dcolumn}
\usepackage{bm}
\usepackage{hyperref}
\usepackage{color}

\begin{document}

\preprint{Phys. Rev. B (accepted)}

\title{Weak antilocalization in topological insulator Bi$_{2}$Te$_{3}$ microflakes}


\author{Shao-Pin Chiu$^1$}
\author{Juhn-Jong Lin$^{1,2,}$}
\email[Electronic address: ]{jjlin@mail.nctu.edu.tw}

\affiliation{$^1$Institute of Physics, National Chiao Tung University, Hsinchu 30010, Taiwan\\
$^2$Department of Electrophysics, National Chiao Tung University, Hsinchu 30010, Taiwan}

\date{\today}

\begin{abstract}

We have studied the carrier transport in two topological insulator (TI) Bi$_{2}$Te$_{3}$ microflakes between 0.3 and 10 K and under applied backgate voltages ($V_{\rm BG}$). Logarithmic temperature dependent resistance corrections due to the two-dimensional electron-electron interaction effect in the presence of weak disorder were observed. The extracted Coulomb screening parameter is negative, which is in accord with the situation of strong spin-orbit scattering as is inherited in the TI materials. In particular, positive magnetoresistances (MRs) in the two-dimensional weak-antilocalization (WAL) effect were measured in low magnetic fields, which can be satisfactorily described by a multichannel-conduction model. Both at low temperatures of $T < 1$ K and under high positive $V_{\rm BG}$, signatures of the presence of two coherent conduction channels were observed, as indicated by an increase by a factor of $\approx$ 2 in the prefactor which characterizes the WAL MR magnitude. Our results are discussed in terms of the (likely) existence of the Dirac fermion surface states, in addition to the bulk states, in the three-dimensional TI Bi$_2$Te$_3$ material.

\end{abstract}


\maketitle

\section{Introduction}

Topological insulators (TIs) belong to a new class of quantum matters, where unique gapless surface states with linear energy-momentum dispersion characteristics (i.e., Dirac fermion states) coexist with gapped bulk states. \cite{Hasan10,Moore10,Qi12} As a consequence of the inherit strong spin-orbit coupling, the spin orientations of the helical surface states are transversely locked to their translational crystal momenta. In the absence of any magnetic scattering, the time-reversal coherent backscattering would thus be suppressed, leading to the well-known weak antilocalization (WAL) effect which manifests positive magnetoresistances (MRs) in low perpendicular magnetic fields. \cite{Bergmann84,Hikami80,McCann06,Tkachov11}

Although an observation of the two-dimensional (2D) WAL effect can be a signature of the existence of TI surface states, the strong spin-orbit coupling in the bulk conduction channel can also contribute to a WAL effect. This situation often happens in the three-dimensional (3D) TIs, such as the p-type Bi$_{2}$Te$_{3}$ and the n-type Bi$_{2}$Se$_{3}$, owing to the high levels of unintentional doping which readily occurs during the sample growth as well as the device fabrication processes. Therefore, the MR data of 3D TIs have often been analyzed in terms of a multichannel-conduction model which considers the potential contributions from both the surface and the bulk states. \cite{Takagaki12,Steinberg11,Chen11,Chen10,Liu11,He11,Checkelsky11} Due to the small surface-to-volume ratios in real samples, a clear-cut separation of the possible surface contribution from the overall carrier transport has remained nontrivial. Even if a surface contribution were separated, it still poses a very difficult task to associate the contribution with either the top or the bottom surface states, or both. To reach a consensus on this issue will definitely require more experiments employing specifically designed samples (e.g., with both top and bottom gates)  and with improved material qualities. Equally important, a good theoretical understanding of the various issues, such as the coupling mechanisms (e..g, inelastic electron relaxation) between the surface and the bulk states, the electron/hole dephasing processes, the Coulomb interaction effect in the limit of strong spin-orbit scattering, and the band bending effect in individual TI materials, is urgently called for before a quantitative analysis and definitive conclusion about the coherent surface transport could be unambiguously drawn.

In this work we have studied the MRs in two exfoliated Bi$_{2}$Te$_{3}$ microflakes between 0.3 and 10 K and under applied backgate voltages $V_{\rm BG}$. We have found the positive MRs in the WAL effect in small perpendicular magnetic fields $B$. Our results indicate an emergence of two coherent conduction channels as either the temperature is reduced to below 1 K or V$_{\rm BG}$ is increased to be higher than a few tens of volts. That is, the prefactor $\alpha$ [Eq.~(\ref{2DWAL})] which characterizes the WAL MR magnitude increases by a factor of $\approx$ 2 from $\approx$ 0.35 to $\approx$ 0.7. These observations are discussed in terms of the possible (partial) decoupling of the surface states from the bulk states. Moreover, we have observed the 2D electron-electron interaction (EEI) effect in the weakly disordered regime, which caused a logarithmic temperature dependent resistance rise at low temperatures. The extracted Coulomb screening parameter is negative, which faithfully reflects a situation of strong spin-orbit scattering as is inherited in the TI materials.

This paper is organized as follows. Section II contains our experimental method for sample fabrication and electrical-transport measurements. Section III contains our experimental results of resistance and MR as functions of temperature, magnetic field, and backgate voltage. Comparison and analyses based on existing theoretical concepts and predictions are made. Possible limitations on the deduced information based on current theoretical understanding are discussed. Our conclusion is given in Sec. IV.

\section{Experimental method}

Single crystals of Bi$_{2}$Te$_{3}$ were prepared by melting and annealing high-purity Bi$_{2}$Te$_{3}$ and Te powders (99.999\% purity) in a sealed quartz ampoule which was continuously stirred. The temperature was rapidly increased to 1000 $^\circ$C for a few hours, and then slowly reduced to 500 $^\circ$C to allow the crystalline nucleation during a period of five days. A subsequent annealing for another five days was followed to allow the ampoule temperature to reduce slowly from 500 to 420 $^\circ$C. The mixture was then cooled to room temperature. The x-ray powder diffraction study demonstrated the genuine crystalline condition of no. 166 space group as referred to the PDF card 820358.

\begin{table*}[tb]
\caption{\label{table_1}%
Parameters for two Bi$_2$Te$_3$ microflake devices. $t$ is the thickness, $w$ is the width, $L$ is the device length (i.e., the closest distance between the two voltage probes in a four-probe configuration), $\mu$ is the hole mobility, $p$ is the hole concentration, $l$ is the elastic mean free path, and $D$ is the diffusion constant. The values of $\mu$, $p$, $l$ and $D$ are for 10 K.}
\begin{ruledtabular}
\begin{tabular}{ccccccccccccc}
Device & $t$ & $w$ & $L$ & $R(300\,{\rm K})$ & $\rho(300\,{\rm K})$ & $R(10\,{\rm K})$ & $\rho(10\,{\rm K})$  & $\mu$ & $p$ & $l$ & $D$ \\

& (nm) & ($\mu$m) & ($\mu$m) & ($\Omega$) & (m$\Omega$\,cm) & ($\Omega$) & (m$\Omega$\,cm) & (cm$^2$/V\,s) & (cm$^{-3}$) & (nm) & (cm$^2$/s) \\ \hline

BT-15 & 270 & 1.9 & 3.9 & 111 & 1.48 & 35.0 & 0.464 & 1690 & 8.0$\times$$10^{18}$ & 69 & 36 \\
BT-24 & 65 & 2.8 & 1.9 & 424 & 4.06 & 84.0 & 0.805 & 1360 & 5.7$\times$$10^{18}$ & 50 & 23 \\
\end{tabular}
\end{ruledtabular}
\end{table*}

Bi$_2$Te$_3$ microflakes were placed on 300-nm SiO$_2$ capped highly doped n-type Si substrates. Four-probe 20-nm thick Au electrodes on the microflakes were fabricated by photolithography. The inset of Fig.~\ref{fig_1}(a) shows an optical micrograph of the BT-24 device. The thickness $t$ and width $w$ of the microflakes were measured with an atomic force microscope. In the present experiment, the relevant sample length $L$ was taken to be the closest distance between the two voltage probes in a four-probe configuration. This manner of defining the effective sample length could cause an error bar in the extracted sample resistivity (and thus, the sheet resistance $R_\square = \rho /t$) by an amount of at most 10\%, see the Supplemental Material. \cite{supplementary}

MR measurements were performed on an Oxford Heliox $^3$He cryostat with a base temperature of 250 mK and equipped with a 4-T superconducting magnet. The temperature was monitored with calibrated RuO$_2$ and Cernox thermometers. The resistances were measured using a Linear Research LR-700 ac resistance bridge and a Stanford SIM-921 ac resistance bridge. An excitation current of 100 nA was applied to avoid electron heating. That is, the voltage drop along the sample length was $\lesssim k_BT/e$ with this amount of excitation current, where $k_B$ is the Boltzmann constant, and $e$ is the electronic charge. A Keithley Model 2635A dc sourcemeter was utilized to provide backgate voltages. Our device parameters are listed in Table~\ref{table_1}.

\section{Results and discussion}
\subsection{Temperature dependence of sheet resistance}

\begin{figure}[tb]
\includegraphics[scale=0.22]{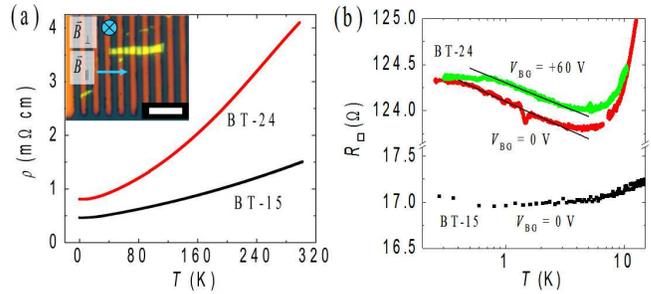}
\caption{\label{fig_1}%
(Color online) (a) Resistivity as a function of temperature for BT-15 and BT-24 devices. Inset: an optical micrograph of the BT-24 device. The scale bar is 10 $\mu$m. (b) Sheet resistance as a function of the logarithm of temperature of BT-15 and BT-24 at low temperatures. The straight lines are least-squares fits to Eq.~(\ref{2DEEI}).}
\end{figure}

The temperature dependence of resistivity $\rho$ of BT-15 and BT-24 reveals overall metallic behavior, as shown in Fig.~\ref{fig_1}(a). A metallic feature of $\rho (T)$ has often been observed in Bi$_{2}$Te$_{3}$ and Bi$_{2}$Se$_{3}$ materials because the unintentional defects (free carriers) are readily generated during the sample growth as well as the device fabrication processes, which cause the Fermi level to shift into the bulk valence or conduction band. In Bi$_{2}$Te$_{3}$, the anti-structural defects in which Bi atoms are found on Te sites  are responsible for the hole doping. \cite{Hor-jpcm10} At liquid-helium temperatures, our two devices show very different $T$ dependences [Fig.~\ref{fig_1}(b)], strongly suggesting a disorder related phenomenon. The sheet resistance $R_\square$ of BT-24 increases with decreasing $T$ below about 4 K. In contrast, the BT-15 device has a sheet resistance of $R_\square$(10\,K) = 17.1 $\Omega$, which is nearly one order of magnitude smaller than that (124 $\Omega$) of BT-24. As a consequence, $R_\square$ remains roughly constant below 6 K in BT-15. The ln\,$T$ increase in BT-24 between 0.4 and 4 K is due to the 2D EEI effect in the weakly disordered regime. Similar logarithmic temperature dependence of $R_\square$ has been seen in several TI samples. \cite{Liu11,Wang11,Takagaki12} The EEI correction to the residual resistance in a quasi-2D conductor is given by \cite{Altshuler-prl80,Lin-prb87a}
\begin{equation}
\frac{\triangle R _\square (T)}{R_\square (T_0)} = - \frac{e^2}{2 \pi^2 \hbar} \left( 1 - \frac34 F \right) R_\square \ln \left( {\frac{T}{T_0} } \right) \,,
\label{2DEEI}
\end{equation}
where $\triangle R_\square (T) = R_\square (T) - R_\square (T_0)$, $2\pi\hbar$ is the Planck constant, \textit{F} is an electron screening factor, and \textit{T}$_{0}$ is a reference temperature (taken to be 4 K in this work). From least-squares fits to Eq.~(\ref{2DEEI}), we obtain 1$-$3\textit{F}/4 = 1.33 and \textit{F} = $-$0.43, while the EEI theory in its simplest form predicts $0 \lesssim F \lesssim 1$. \cite{F-tilde} Recently, a negative $F$ value has also been extracted by Takagaki {\it et al.} \cite{Takagaki12} in Cu-doped Bi$_{2}$Se$_{3}$, but they did not connect their observation and discussion in terms of the strong spin-orbit coupling property of the TI materials.  In the original theory of Altshuler {\it et al.}, \cite{Altshuler-ssc82} the triplet term in the EEI effect in the diffusion channel is found to be suppressed by strong spin-orbit coupling. However, the theory does not yield a negative $F$ value. Empirically, Wu {\it et al.}  \cite{Wu-prb95} had previously demonstrated, by using a series of TiAl alloys doped with dilute heavy Au atoms, that sufficiently stronger spin-orbit coupling can cause a more negative $F$ value. Whether this is also the origin in the TI materials deserves further theoretical clarification. \cite{F-value} Here we also would like to point out that, due to the finite width of our voltage leads as compared with the relevant sample length in a four-probe configuration, our evaluation of $R_\square$ could be overestimated by an amount of at most 10\% (see the Supplemental Material \cite{supplementary}). Taking the largest possible 10\% uncertainty in $R_\square$ [which appears on the right-hand side of Eq.~(\ref{2DEEI})] into account, our extracted $F$ value can be recalculated to be $F \simeq$ $-$0.52$\pm$0.09.

Figure~\ref{fig_1}(b) also plots the $R_\square (T)$ of BT-24 under an applied backgate voltage of \textit{V}$_{\rm BG}$ = $+$60 V. The overall $R_\square (T)$ curve is slightly higher than the corresponding curve under zero backgate voltage, indicating hole doping. The straight line is a least-squares fit to Eq.~(\ref{2DEEI}). It indicates a coefficient of 1$-$3\textit{F}/4 = 1.23 and \textit{F} = $-$0.31. (If taking a possible 10\% overestimate in the $R_\square$ value into account, our $F$ value would read $ F = -$0.39$\pm$0.08.) This result illustrates that the EEI effect persists in Bi$_2$Te$_3$ when a large positive \textit{V}$_{\rm BG}$ was applied and the surface and bulk states became (partly) decoupled. (The possible decoupling under a $V_{\rm BG}$ = $+$60 V is asserted through the measurements of the MR dips in the WAL effect, see below.) On the other hand, in Eq.~(\ref{2DEEI}), since the magnitude of the resistance correction due to the 2D EEI effect scales linearly with $R_\square$, a ln\,$T$ increase in $R_\square$ thus must be minute in BT-15, as mentioned.

In the above discussion, we have ignored the WAL correction to $R_\square (T)$. Theoretically, the WAL effect is known to cause an opposite decrease of $R_\square$ with reducing $T$, and is given by \cite{Lin-prb87a,Abrahams-prl79} $\triangle R_\square (T) / R_\square (T_0) = -\alpha \tilde{p} (e^2/2 \pi^2 \hbar) R_\square {\rm ln} (T/T_0)$, where $\alpha$ is defined in Eq.~(\ref{2DWAL}), and $\tilde{p}$ is the exponent of temperature in the electron dephasing time $\tau_\varphi \propto T^{- \tilde{p}}$. In our case, this contribution does not seem to become important until $T$ is lowered to subkelvin temperatures. Indeed, the seemingly saturation of $R_\square$ below $\sim$ 0.4 K in BT-24 under zero backgate voltage is not due to the Joule heating (as was discussed in the Experimental Method), it is most likely a signature of the onset of the WAL effect. This interpretation is supported by the fact that an onset of the downward deviation from the ln\,$T$ dependence occurs at a slightly higher $T \approx$ 0.7 K when a $V_{\rm BG}$ = $+$60 V was applied, which caused a slightly more negative $\alpha$ value [Fig.~\ref{fig_4}(b)], and hence a slightly larger WAL contribution. \cite{Bergmann84, Lin-prb87a} We may roughly estimate that the EEI term is $\sim$ 5 times greater than the WAL term, i.e., $(1 - 3F/4)/|\alpha \tilde{p}| \approx 5$, where we have used our experimental values of $\alpha \sim - 0.5$ and $\tilde{p} \approx$ 0.5 (see Sec. III B). In short, the observation of increasing $R_\square$ with decreasing $T$ in Fig.~\ref{fig_1}(b) confirms the important role played by the 2D EEI effect in the presence of both weak disorder and strong spin-orbit coupling in Bi$_2$Te$_3$. \cite{strictly} To the best of our knowledge, the connection of negative $\tilde{F}$ values with strong spin-orbit coupling in TIs has not been pointed out in any previous theoretical and experimental studies.

\begin{figure}[tb]
\includegraphics[scale=0.21]{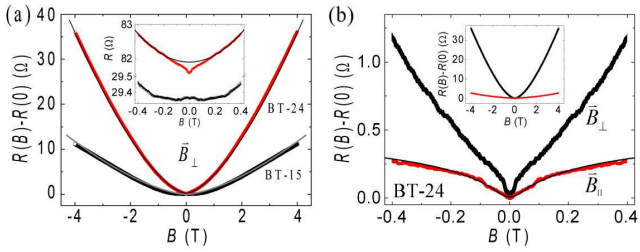}
\caption{\label{fig_2}%
(Color online) (a) MR curves of BT-15 and BT-24 in perpendicular \textit{B} field at $T$ = 0.3 K. The solid curve drawn through BT-15 (BT-24) describes a $|B|^\gamma$ power law, with $\gamma$ = 1.53 (1.55). The inset shows the low magnetic field regime of $|B| \leq$ 0.4 T. The solid curve drawn through BT-24 describes a $B^2$ power law. (b) MR curves of BT-24 in perpendicular and parallel \textit{B} fields at $T$ = 0.3 K. The solid curve drawn through the parallel MR data is the 2D WAL theoretical prediction with an electron dephasing length $L_\varphi \simeq$ 420 nm (see Refs. \onlinecite{ng-prb93} and \onlinecite{Chiu-unpublished}). The inset shows the large magnetic field regime of $|B| \leq$ 4 T.}
\end{figure}

Figure~\ref{fig_2}(a) shows the MR data of the BT-15 and BT-24 devices at 0.3 K with the $B$ field applied perpendicular to the microflake plane. Note that the MR (ignoring the dip around $B$ = 0) reveals an approximate \textit{B}$^{2}$ dependence in the low magnetic field regime of $|B| \lesssim 1$ T, see the inset of Fig.~\ref{fig_2}(a). This may be tentatively ascribed to the classical MR and approximated by: \cite{MR} $[R(B)-R(0)]/R(0) \simeq (\mu B)^{2}$. We thus obtain the mobility $\mu \approx$ 1690 (1360) cm$^{2}$/V\,s in BT-15 (BT-24). The hole concentration $p = 1/(\rho |e| \mu)$ is then calculated to be $\approx 8.0 \times 10^{18}$ ($\approx 5.7 \times 10^{18}$) cm$^{-3}$ in BT-15 (BT-24). \cite{Hall} These values of $\mu$ and $p$ are in line with those values deduced for the Bi$_2$Te$_3$ samples with compatible resistivities. \cite{He11} Using the free-electron model, we estimate the thermal diffusion length $L_T = \sqrt{D \hbar/k_BT} >$ 65 nm at $T \lesssim$ 4 K in BT-24 ($D$ being the diffusion constant). Therefore, the 2D EEI effect may occur at low $T$ in this device, as we have seen above.

Furthermore, the overall feature of the MR curves of BT-15 and BT-24 in the wide magnetic field range $|B| \leq$ 4 T are nearly $T$ independent between 0.3 and 10 K and can be satisfactorily described by the power law $[R(B) - R(0)] \propto |B|^\gamma$, with $\gamma$ = 1.53$\pm$0.01 (1.55$\pm$0.01) for BT-15 (BT-24), see the solid curves drawn through the MR data in the main panel of Fig.~\ref{fig_2}(a). The authors of Refs. \onlinecite{Takagaki12} and \onlinecite{Takagaki07} proposed that a magnitude of an intermediate power of $1 < \gamma < 2$ would be suggestive of the presence of multiple types of carriers.

In the inset of Fig.~\ref{fig_2}(a) we plot the MR curves measured in the low field regime of $|B| \leq$ 0.4 T and at $T$ = 0.3 K. A resistance dip in the MR curve of BT-24 is clearly seen, manifesting the WAL effect. On the other hand, the WAL effect in BT-15 is obscured by the relatively large universal conductance fluctuations (UCFs). \cite{Lee87,Matsuo12,Yang-prb12} In the rest of this paper, we shall thus focus our analysis of the WAL effect only on the BT-24 sample. It is, however, worth noting in passing that the UCF signals do allow us to extract the electron dephasing length $L_\varphi (T)$. Our analysis of the root-mean-square UCF magnitudes at various temperatures led to $L_\varphi$ values which are consistent to within $\sim$ 40\% with those corresponding values deduced from the WAL method, see Fig.~\ref{fig_3}(d).

\subsection{Weak-antilocalization magnetoresistance: Temperature dependence}

In Fig.~\ref{fig_2}(b), we plot the MR curves of BT-24 at $T$ = 0.3 K and with the \textit{B} field applied either perpendicular to the microflake plane or parallel to the microflake plane and in the direction of the current flow. In both $B$ field orientations, the MR dips around $B$ = 0 are evident. Note that in the parallel $B$ field orientation, only the bulk states can possibly contribute to the quantum-interference correction to the classical parabolic MR. Therefore, an observed MR dip in this case unambiguously indicates that the bulk states must lie in the strong spin-orbit scattering limit, giving rise to the WAL effect. This observation does not support the recent theoretical prediction of a weak-localization effect (which should manifest a MR ``peak") in the Bi$_2$Te$_3$ material. \cite{Lu-prb11} Also, we have carried out least-squares fits of the parallel MR curve to the pertinent 2D WAL theoretical prediction \cite{ng-prb93,Chiu-unpublished} [the solid curve in Fig.~\ref{fig_2}(b)] and obtained an electron dephasing length of $L_\varphi$(0.3\,K) $\approx$ 420 nm. This length scale is larger than the thickness (65 nm) of the BT-24 microflake, and hence our bulk channel must be 2D with regard to the WAL effect. \cite{He-2D} Thus, in the case of perpendicular $B$ field orientation, both the surface and bulk states would contribute to the 2D WAL effect and the measured MR curves have to be analyzed in terms of a multichannel-conduction model.

\begin{figure}[tb]
\includegraphics[scale=0.2]{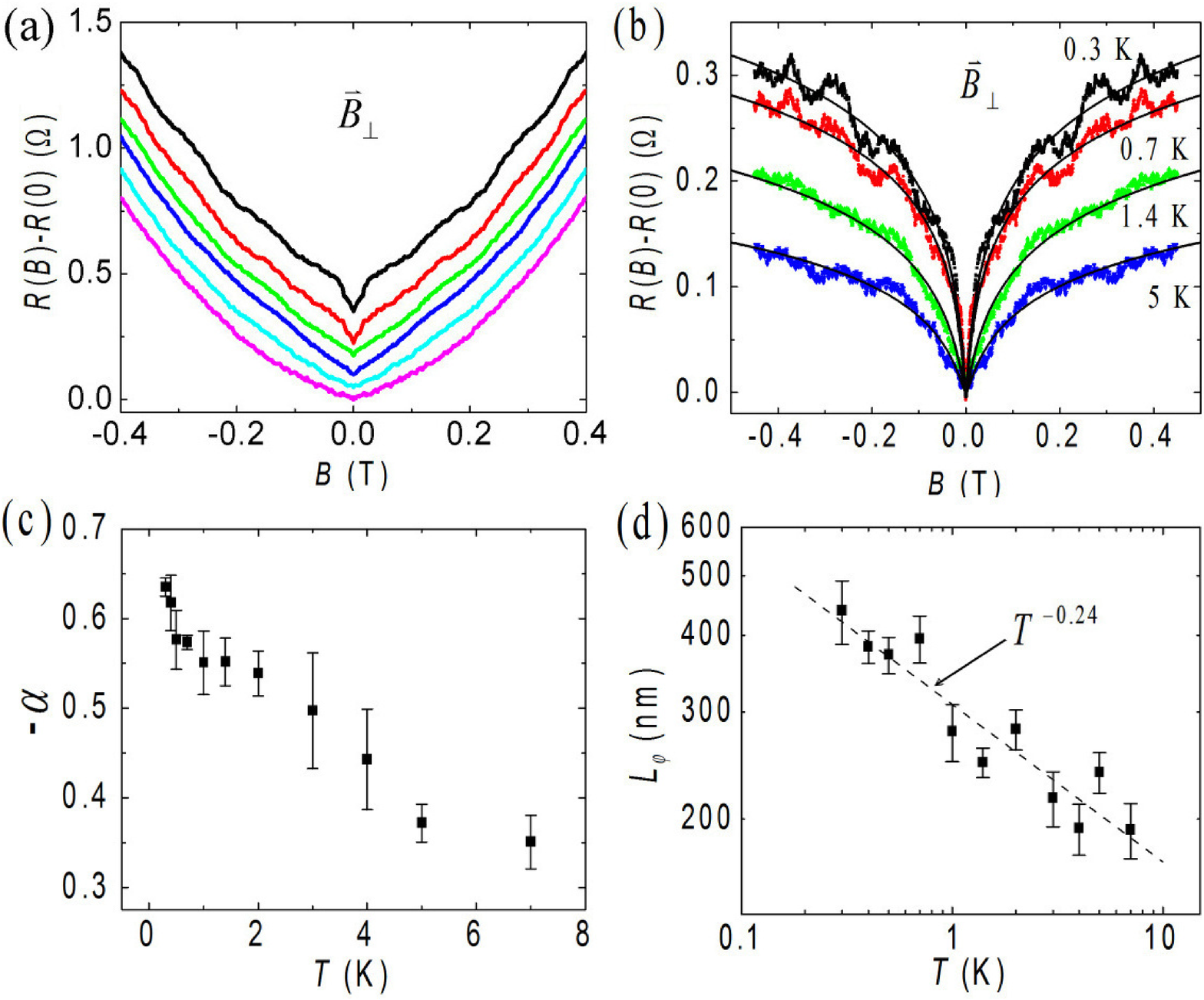}
\caption{\label{fig_3}%
(Color online) BT-24 microflake in perpendicular $B$ fields and under $V_{\rm BG}$ = 0. (a) MR curves at several $T$ values (from top down): 0.30, 0.50, 1.0, 2.0, 5.0, and 10 K. The curves are vertically offset for clarity. (b) MR curves after subtracting away the 10-K background curve and at four representative $T$ values, as indicated. The solid curves are the theoretical predictions of Eq.~(\ref{2DWAL}). Note that universal conductance fluctuations are visible. (c) Variation of extracted parameter $- \alpha$ with temperature. As $T$ decreases from 10 to 0.3 K, $- \alpha$ increases by a factor of $\approx$ 2. (d) Variation of extracted electron dephasing length \textit{L}$_{\varphi}$ with temperature. The straight dashed line is drawn proportional to $T^{-0.24}$ and is a guide to the eye.}
\end{figure}

In Fig.~\ref{fig_3}(a), we show the MR curves of BT-24 measured in the low, perpendicular magnetic field regime of $|B| \leq$ 0.4 T and at several $T$ values between 0.3 and 10 K, as indicated in the caption to Fig.~\ref{fig_3}. Note that the curves are vertically offset for clarity. \cite{offset} It can be seen that the MR dips increases with decreasing $T$, suggesting the more pronounced quantum-interference carrier transport at lower $T$. As $T$ increases, the MR dip eventually diminishes around $T \sim$ 10 K. Therefore, we take this 10-K MR curve as the background curve and subtract it away from those MR curves measured at lower $T$ values. This procedure gives rise to those WAL MR curves shown in Fig.~\ref{fig_3}(b) (for clarity, only four representative curves are plotted). The 2D WAL MR in perpendicular $B$ fields and in the limit of strong spin-orbit scattering (which is pertinent to TIs) can be written as \cite{Hikami80,McCann06,Tkachov11,Lin-prb87b}
\begin{equation}
\frac{\triangle R_\square  (B)}{[R_\square (0)]^2} = - \alpha \frac{e^2}{2\pi ^2 \hbar} \left[ \Psi \left( \frac{1}{2} + \frac{B_\varphi} {B} \right) - \ln \left( \frac{B_\varphi}{B} \right) \right] \,,
\label{2DWAL}
\end{equation}
where $\triangle R_\square (B) = R_\square(B) - R_\square (0)$, $\Psi (x)$ is the digamma function, the characteristic magnetic field $B_\varphi = \hbar /(4eL_\varphi^2)$, and $L_\varphi = \sqrt{D \tau_\varphi}$. $\alpha$ is a parameter whose value reflects the number of conduction channels. For the 2D surface states in a 3D TI, $\alpha = - 1/2$ for a single coherent topological surface channel, and $\alpha = - 1$ for two independent coherent transport channels with similar $L_\varphi$'s. In practice, for yet to be identified reasons, the experimentally extracted $\alpha$ values often differ from these two ideal values. For simplicity and also in order to reduce the number of adjustable parameters, in this work we take $L_\varphi$ as an effective dephasing length averaged over the channels. \cite{average} The main objective of this paper is thus to show that the value of $\alpha$ is indeed tunable, which to a good extent seems to signify the existence of conductive surface states.

Our measured MR data plotted in Fig.~\ref{fig_3}(b) can be well described by the predictions of Eq.~(\ref{2DWAL}) (the solid curves). The extracted $\alpha (T)$ and $L_{\varphi} (T)$ values are plotted in Figs.~\ref{fig_3}(c) and \ref{fig_3}(d), respectively. Our experimental value of $- \alpha$ monotonically increases from $\simeq$ 0.35 to $\simeq$ 0.64 as $T$ is reduced from 10 to 0.3 K. (If taking a possible 10\% uncertainty in the $R_\square$ value into account, our extracted $-$$\alpha$ value would be slightly modified to vary from 0.39 to 0.70.) This variation of $\alpha$ with $T$ is well beyond our experimental uncertainties, and it is plausible to think that such a change in the $\alpha$ value by a factor of $\sim$ 2 does reflect a decoupling (doubling) of the charge transport channels at sufficiently low $T$. Note that several recent studies of Bi$_{2}$Se$_{3}$ have also reported temperature dependent $\alpha$ values which lie between $-$0.3 and $-$0.6, but not between $-$0.5 and $-$1. \cite{Liu11,Chen11,Chen10,Steinberg11,He11,Checkelsky11} For the Bi$_2$Te$_3$ material, to our knowledge, the temperature behavior of $\alpha$ has not yet been reported in the literature. He {\it et al.} have recently reported an $\alpha$ = $-$0.39 at $T$ = 2 K, \cite{He11} while our value is $\alpha$(2\,K) $\simeq$ $-$0.54. In our case, an $- \alpha$(0.3\,K) value smaller than unity might imply that the decoupling is incomplete (partly due to the high carrier density in BT-24). The exact reason why ultralow temperature facilitates decoupling should await further theoretical investigations. Tentatively, if there should exist an inelastic relaxation process between the surface carriers and the bulk low-lying excitations (e.g., surface-carrier--bulk-phonon scattering \cite{Sebastien11}), the scattering strength would decrease with reducing $T$. Then, a higher degree of decoupling of the surface and bulk states could likely take place at lower $T$. \cite{phonon} For comparison, we would like to stress that in the conventional metals and alloys with strong spin-orbit scattering, such as Au-doped Cu (Ref. \onlinecite{Huang-prl07}) and AuPd (Refs. \onlinecite{Lin-prb87b} and \onlinecite{Zhong-prl10}) thin films, the MR dips in 2D WAL effect have been firmly observed, where one {\em always} finds a constant $\alpha$ = $-$1/2 in the wide $T$ range from subkelvin temperatures up to above 20 K.

Our extracted \textit{L}$_{\varphi}$ values of BT-24 are plotted in Fig.~\ref{fig_3}(d). As $T$ decreases from 10 to 0.3 K, $L_\varphi$ increases from $\simeq$ 200 to $\simeq$ 440 nm. The straight dashed line is drawn proportional to $L_\varphi \propto T^{-0.24}$ and is a guide to the eye. This slope corresponds to an effective exponent of temperature $\tilde{p} \simeq$ 0.48 in $\tau_\varphi \propto T^{-\tilde{p}}$, which is considerably lower than that ($\tilde{p}$ = 1) would be expected for the quasielastic Nyquist electron-electron scattering process in 2D. \cite{Altshuler82,Wu-prb12} This suggests the existence of additional weakly $T$ dependent magnetic or nonmagnetic electron dephasing processes which are noticeable in this device over our measurement temperature range. \cite{Huang-prl07,Lin02} Our extracted $L_\varphi$ values are larger than the thickness of the microflake, justifying the 2D WAL characteristics in our samples. Furthermore, our value of $L_\varphi$(0.3\,K) inferred from the perpendicular MR, Eq.~(\ref{2DWAL}), is in good agreement with that ($\simeq$ 420 nm) deduced from the parallel MR, see Fig.~\ref{fig_2}(b) and Ref. \onlinecite{Chiu-unpublished}. This close agreement in the extracted $L_\varphi (B_\perp)$ and $L_\varphi (B_\parallel)$ values strongly indicates the self-consistency of our experimental method and data analysis as well as the reasonable validity of applying the WAL theory in its current form \cite{Hikami80} to the TI Bi$_2$Te$_3$ material. Whether any modifications might need to be incorporated into Eq.~(\ref{2DWAL}) to take into account any subtle effect(s) that might arise from the multichannel feature and the specific materials properties of Bi$_2$Te$_3$ have to await future theoretical investigations. For instance, it would be of interest and importance to see if the frequent experimental observations of an $|\alpha|$ value smaller than 0.5 could be answered.

\subsection{Weak-antilocalization magnetoresistance: Backgate-voltage dependence}

\begin{figure}[tb]
\includegraphics[scale=0.22]{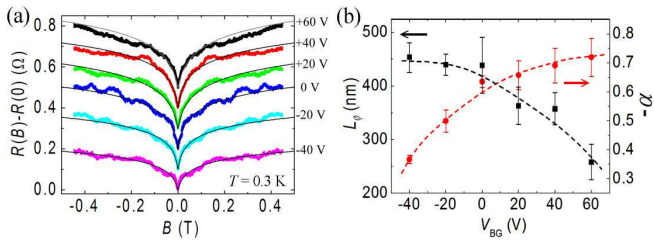}
\caption{\label{fig_4}%
(Color online) BT-24 microflake at $T$ = 0.3 K. (a) MR data measured in perpendicular \textit{B} field and under several \textit{V}$_{\rm BG}$ values, as indicated. The data are vertically offset for clarity. The solid curves are least-squares fits to Eq.~(\ref{2DWAL}). Note that resistance fluctuations are visible. (b) Variation of the extracted parameter $- \alpha$ and electron dephasing length \textit{L}$_{\varphi}$ with $V_{\rm BG}$. The dashed curves are guides to the eye.}
\end{figure}

We have further performed the MR measurements under different applied backgate voltages $V_{\rm BG}$ and at $T$ = 0.3 K. Figure~\ref{fig_4}(a) clearly shows that the size of the MR dip increases monotonically as \textit{V}$_{\rm BG}$ progressively increases from $-$40 to $+$60 V. From least-squares fits to Eg.~(\ref{2DWAL}), we have obtained the values of $\alpha$ and \textit{L}$_{\varphi}$ under various \textit{V}$_{\rm BG}$ values. Figure~\ref{fig_4}(b) reveals that, as \textit{V}$_{\rm BG}$ increases from $-$40 to $+$60 V, $- \alpha$ monotonically increases from $\simeq$ 0.36 to $\simeq$ 0.72, while \textit{L}$_{\varphi}$ monotonically decreases from $\simeq$ 450 to $\simeq$ 250 nm. (If taking a possible 10\% overestimate in the $R_\square$ value into account, our $-$$\alpha$ value would be slightly modified to vary from 0.40 to 0.79, while the extracted $L_\varphi$ values are barely affected.) These amounts of variation in $\alpha$ and $L_\varphi$ are remarkable, since a change of $V_{\rm BG}$ from $-$40 V to $+$60 V only causes a $\simeq$ 2\% reduction in the hole concentration $p$ in this device. [Such a minor change in $p$ can be ascribed to the facts (1) that our applied $V_{\rm BG}$ partly dropped in the highly doped n-type Si substrate, (2) that the relatively thick SiO$_2$ layer reduced the ability to electrostatically modulate the interface carrier density, and (3) that our sample possessed a relatively high $p$ level. Recall that, in Fig.~\ref{fig_1}(b), the $R_\square$ value of BT-24 increased only by a tiny amount of $\lesssim$ 0.2\% as $V_{\rm BG}$ was varied from 0 to $+$60 V.] For comparison, we would like to point out that several previous experiments using the n-type Bi$_{2}$Se$_{3}$ have also found that, as the electron density ($n$) is lowered by a sufficiently large negative $V_{\rm BG}$, the value of $- \alpha$ increases (i.e., $\alpha$ becomes more negative) while the value of $L_{\varphi}$ decreases. \cite{Chen11,Chen10,Steinberg11,Checkelsky11} Since the $p$/$n$ carriers in the bulk channel are (partly) depleted away under a large positive/negative $V_{\rm BG}$, a decoupling of the two channels thus could possibly take place. This in turn is reflected by the increase in the $-$$\alpha$ value by a factor of $\approx$ 2.

Because our extracted $- \alpha$ value is always smaller than unity, we have presumed that there exists only one conductive surface channel in BT-24. This presumption is in accord with a recent report by Takagaki {\it et al.} \cite{Takagaki12} who have pointed out that the seemingly absence of the WAL effect for one of the surfaces was a common finding in a number of studies of 3D TIs. \cite{Liu11,Chen11,Chen10,Steinberg11,He11,Wang11} In our case, the top surface layer might have been deteriorated owing to its constant exposure to the air atmosphere as well as its undergoing the device fabrication process. Furthermore, our BT-24 is 65 nm thick and highly doped, and hence the top surface state, if any exists, would be too far away to be tuned by $V_{\rm BG}$. Therefore, we tentatively ascribe the observed conductive surface states to be associated with the bottom surface of the microflake. It is also important to note that we have ascribed the additional coherent transport channel to the TI bottom states, rather than to an electron accumulation layer. To rule out the latter scenario, we have measured the low-$T$ resistance as a function of backgate voltage in a BT-23 microflake which possessed very similar electronic properties to those of the BT-24 microflake. Our measured resistance curve $R(V_{\rm BG})$ at every temperature always revealed a maximum at a characteristic backgate voltage $V_{\rm BG}^\ast$, which signified that the Fermi energy was indeed shifted to become aligned with the Dirac point at this particular backgate voltage (see further discussion in the Supplemental Material \cite{supplementary}).

Finally, the reason for the marked effect of $V_{\rm BG}$ on $L_\varphi$ [Fig.~\ref{fig_4}(b)] is theoretically less clear. Under a large positive $V_{\rm BG}$, the carriers are slightly depleted away from the bottom surface. The depletion would certainly lead to deteriorated Coulomb screening between the carriers, leading to an enhanced electron-electron (hole-hole) scattering rate and thus a shortened $L_\varphi$. However, as mentioned, the hole concentration is reduced only by 2\% under a $V_{\rm BG}$ = $+$60 V in this BT-24 device. It is thus conjectured whether such a small reduction in $p$ could have already caused a notable change in $L_\varphi$. There could be other explanations, for example, based on a change in the degree of coupling between bulk and surface states, which would take place in the presence of large backgate voltages. A quantitative account for the suppression of $L_\varphi$ by $V_{\rm BG}$ must await further theoretical investigations.

\section{Conclusion}

We have measured the resistances and magnetoresistances of two Bi$_{2}$Te$_{3}$ microflakes at low temperatures and under applied backgate voltages. Low-temperature resistance corrections due to the 2D EEI effect in the presence of weak disorder are observed. The extracted Coulomb screening parameter $F$ is negative, which is in line with the situation of strong spin-orbit scattering as is inherited in the TI materials. Positive MR dips in small perpendicular magnetic fields are measured, which can be satisfactorily described by the existing 2D WAL theory by taking into account a multichannel-conduction model in a TI material. Our extracted $\alpha (T,V_{\rm BG})$ values, which increase by a factor of $\approx$ 2 from $\approx$ 0.35 to $\approx$ 0.7, suggest the emergence of two coherent conduction channels as either the temperature decreases to subkelvin temperatures or the backgate voltage increases to several tens of volts. This doubling of the conduction channels seems to signify the (partial) decoupling of the surface and the bulk states. In short, our overall results provide qualitative, but not yet quantitative, evidence for the likely existence of Dirac fermion states in the 3D Bi$_2$Te$_3$ material. To achieve a clear-cut separation of the contributions from the individual surface and bulk states undoubtedly would require further theoretical formulations and experimental measurements.

\begin{acknowledgments}

The authors thank Zhao-guo Li, Tai-shi Chen and Feng-qi Song for providing us with the samples used in this study and for their reading through an early version of the manuscript. This work was supported by the Taiwan National Science Council through Grant No. NSC 100-2120-M-009-008 and the MOE ATU Program.

\end{acknowledgments}


\begin{thebibliography}{00}
\bibitem{F-tilde} In the formal EEI theory, the screening parameter $F$ should be replaced by the parameter $\tilde{F} = (32/3F) [(1+F/2)^{3/2} -1 - 3F/4]$. However, $\tilde{F}$ equals $F$ to within $\approx$ 10\% for $-1 \lesssim F \lesssim 1$. It is thus not critical to distinguish between $\tilde{F}$ and $F$ in most experiments. See, for example, Ref. \onlinecite{Wu-prb95}.
\bibitem{strictly} Strictly speaking, the above results do not allow us to separate the individual contributions of the surface and the bulk states.
\bibitem{Hall} The $p$ value determined this way is a few times smaller than that inferred from the Hall measurement on the BT-15 device. This difference is not critical for the present discussion and does not affect the conclusion drawn in this work.
\bibitem{He-2D} In Ref. \onlinecite{He11}, the authors had studied a 50-nm thick Be$_2$Te$_3$ film at $T$ = 2 K. They ascribed their measured MR dip in parallel $B$ fields to the 3D WAL effect but extracted a comparatively large value of $L_\varphi$ = 331 nm. Then, their bulk channel must be 2D, instead of 3D, with regard to the WAL effect.
\bibitem{offset} Without vertical offsets, our measured MR curves (except the dips) at all temperatures below 10 K essentially overlap, due to the metallic feature of the sample.
\bibitem{average} In practice, the electron dephasing lengths of the surface and the bulk channels can not be unambiguously distinguished. Nevertheless, according to the current understanding (see Ref. \onlinecite{Steinberg11}), the two length scales [$(L_\varphi)_{\rm surface}$ and $(L_\varphi)_{\rm bulk}$] differ by no more than a factor of $\sim$ 2. This in turn could cause an error bar of $\lesssim$ 40\% in the difference, if any exists, between  the effective $L_\varphi$ and the surface $(L_\varphi)_{\rm surface}$ values.
\end{thebibliography}
\end{document}